# Exchange bias in GeMn nanocolumns: the role of surface oxidation

**S. Tardif[1,2], S. Cherifi[1,3(a)], M. Jamet[2], T. Devillers[2], A. Barski[2], D. Schmitz[4], N. Darowski[4], P. Thakur[5], J. C. Cezar[5], N. B. Brookes[5], R. Mattana[6] and J. Cibert[1]**

[1] Institut Néel, CNRS and UJF, 25 rue des Martyrs, BP166, F-38042 Grenoble, France
[2] INAC, CEA-Grenoble, 17 rue des Martyrs, F-38054 Grenoble, France
[3] IPCMS, CNRS and UdS, 23 rue du Loess, BP43, F-67034 Strasbourg, France
[4] Helmholtz Centre Berlin, Albert-Einstein-Str. 15, D-12489 Berlin, Germany
[5] European Synchrotron Radiation Facility, BP 220, F-38043 Grenoble, France
[6] Unité Mixte de Physique CNRS-Thales, Route départementale 128, F-91767 Palaiseau, France

**Abstract**

We report on the exchange biasing of self-assembled ferromagnetic GeMn nanocolumns by GeMn-oxide caps. The x-ray absorption spectroscopy analysis of this surface oxide shows a multiplet fine structure that is typical of the $Mn^{2+}$ valence state in MnO. A magnetization hysteresis shift $|H_E| \sim 100$ Oe and a coercivity enhancement $\Delta H_c \sim 70$ Oe have been obtained upon cooling (300-5K) in a magnetic field as low as 0.25 T. This exchange bias is attributed to the interface coupling between the ferromagnetic nanocolumns and the antiferromagnetic MnO-like caps. The effect enhancement is achieved by depositing a MnO layer on the GeMn nanocolumns.

Author to whom correspondence should be addressed. Electronic mail: cherifi@ipcms.u-strasbg.fr

The electrical control of the magnetization is presently one of the most explored aspects in modern magnetism. In addition to the understanding of fundamental aspects, this possibility offers a probable direction towards the development of hybrid devices using, e.g. magnetic multiferroics [1], or taking advantage of the carrier-mediated magnetism in magnetic semiconductor based architectures [2]. In fact, the experimental demonstration of a pure electric field manipulation of the magnetization in (Mn,Ga)As [3] shows the outstanding possibility of controlling magnetism in dilute magnetic semiconductors through electrical gates. The prospect of inducing an exchange bias (EB) in magnetic semiconductors [4] adds also an exceptional functionality to these materials. Moreover, the electric current-controlled EB [5] or the electric field-controlled EB [6, 7] show a promising potential for advanced applications such as fully electric-controlled memory devices.

We have recently demonstrated the ferromagnetic behavior of self-organized GeMn nanocolumns surrounded by a quasi-pure Ge matrix [8, 9]. We report here on the possibility of inducing a significant exchange bias in these nanocolumns by controlling their surface oxidation and by following a regular magnetic field-cooling procedure. The magnetic and spectroscopic properties of non-oxidized and partially-oxidized nanocolumns are studied by means of polarized x-ray absorption spectroscopy and superconducting quantum interference device (SQUID) magnetometry.

The GeMn nanocolumns were grown on Ge(001) by using solid sources molecular beam epitaxy. A 30 nm-thick Ge buffer layer was grown first at 250°C on an epi-ready Ge(001) wafer. Subse-quently, 80 nm-thick GeMn films containing a total amount of about 10 % of Mn were grown at 90°C by co-evaporating Ge and Mn at low deposition rates (about 0.15 Å/s). A two-dimensional spinodal decomposition [10] occurring during the growth leads to the formation of Mn-rich nanocolumns. In these growth conditions, self-organized nanocolumns of about 3 nm in diameter with a density of about $3x10^4$ columns/$\mu m^2$ are estimated from transmission electron microscopy (TEM) observations. Fig. 1 shows typical TEM transverse and plane views of the self-assembled GeMn nanocolumns (dark regions). These GeMn nanocolumns are found to be Mn-rich and they are surrounded by a quasi-pure Ge matrix (Mn contents < 1%), as evidenced by local electron energy loss spectroscopy measurements [9].

The magnetic properties of the nanocolumns have been analyzed by means of a SQUID magnetometer and by using x-ray absorption spectroscopy (XAS) and x-ray magnetic circular dichroism (XMCD) measurements at the Mn $2p \rightarrow 3d$ resonance peaks ($L_{2,3}$ edges). The XAS and XMCD experiments have been performed at the electron storage rings BESSY II (beamline UE-46-PGM1) and ESRF (beamline ID08). All the measurements were carried out in the total electron yield detection mode at 5 K in superconducting magnets allowing the application of a magnetic field of 5 T. The field was applied in the direction of propagation of the incident x-rays with the sample oriented at 30 degrees with respect to the incident photon beam.

Three different samples, prepared under the same growth conditions i.e., same columns' size and density (diameter 3 nm, length 80 nm and spacing 6-7 nm), have been studied. The only difference was the surface capping: (i) uncapped-GeMn, leading to 20 nm –thick oxide caps after water treatment for 15 minutes; (ii) Ge-capped GeMn, leading to about 5 nm –thick oxide caps after water treatment for 15 minutes; and (iii) GeSi-capped GeMn, where an amorphous Si(3nm) layer in addition to the Ge(0.5nm) capping allows an efficient protection of the nanocolumns against oxidation. The de-ionized water treatment permits the dissolution of the native germanium dioxide [11] and induces a local oxidation of the Mn-rich nanocolumns. The oxidation of Mn:Ge upon oxygen incorporation and the formation of Mn-O-Mn complexes has been predicted to decrease the system net magnetization [12]. In the GeMn nanocolumns, the magnetic signal measured by SQUID magnetometry before water treatment has to be multiplied by a numerical factor (< 1) to fit the magnetic signal after water cleaning, while the magnetization remains unaltered in the well protected (SiGecapped) samples. The indirect estimation of the manganese oxide amount by measuring the decay of the magnetic signal upon oxidation is found to be most suitable for this system where the oxidized Mn caps represents only a small fraction of the 10%Mn atoms that are distributed in the GeMn layer.

The x-ray absorption spectra measured at the Mn $L_{2,3}$ edges in the well-protected GeMn nanocolumns show single broad peaks (FWHM = 5.7 eV) related to the metallic character of the GeMn nanocolumns and to the probable hybridization with Ge. Upon oxidation, the XAS and XMCD line shapes become strikingly different as they develop a multiplet fine structure (Fig. 2). In the oxidized nanocolumns, the Mn $L_3$ absorption edge displays several peaks labeled A (640.3 eV), B (641.5 eV) and C (644.0 eV) and the Mn $L_2$ edge displays two peaks labeled D (651.0 eV) and E (652.4 eV). The XAS multiplet structure and the spacing between A-B(1.2 eV) and A-C(3.7 eV) are typical for the $Mn^{2+}$ valence state in MnO as discussed by Gilbert et al. [13] and observed in different XAS measurements [14,15]. Although the oxidation of the Mn-rich GeMn nanocolumns surrounded by a quasi-pure Ge matrix could form complex (Ge,Mn)O oxides, the gradual formation of the typical $Mn^{2+}$ multiplet fine structure upon oxidation is the signature of manganosite-like (i.e., a MnO-like oxide) caps forming at the surface the GeMn nanocolumns.

In the well-protected non-oxidized nanocolumns, the magnetic field-dependent intensity of the XMCD signal at the Mn $L_3$ peak closely follows the magnetization loop M(H) measured by SQUID magnetometry (fig. 3(a)), suggesting that the topmost layers (about 5-7 nm) probed by XMCD are indeed representative of the properties of the whole GeMn nanocolumns' length. When the columns are oxidized over 20 nm (fig. 3(b)), a clear difference in the field-dependent signal measured by the two methods is observed. This is due to the in-depth sensitivity of the SQUID magnetometer that allows us to probe the whole film thickness –i.e., the oxidized caps as well as the ferromagnetic GeMn nanocolumns underneath– versus the surface sensitivity of soft x-ray magnetic dichroism that mostly probes the oxidized caps.

We have exploited these GeMn-oxide caps (20 nm) to induce exchange bias with the base ferromagnetic column layer (60 nm) following the well-established field-cooling procedure (FC) below the MnO blocking temperature (118 K). The detailed ferromagnetic properties of the non-oxidized GeMn nanocolumns could be found in ref. [8,9]. To precisely determine any modification in the magnetization hysteresis loops (shift or coercive field increase), M(H) measurements have been performed before and after FC from 300 K to 5 K at different magnetic field values applied in the film plane (i.e., perpendicular to the columns).

As shown in Fig. 4 and Fig. 5, a positive cooling field shifts the magnetization curve towards the negative field values ($H_E$ negative) while cooling the system under a negative magnetic field shifts the magnetization towards the positive field values ($H_E$ positive). The shift of the hysteresis loop upon the FC procedure is accompanied by an enhancement of the coercive field. These two effects are the signature of a clear EB. The same result was obtained in different GeMn samples where the exchange field was in the range of 65-100 Oe. Only the coercive field varied from 30-100 Oe, depending on the oxide layer thickness. Our study revealed in addition that cooling under a field as low as 0.25 T is sufficient to induce EB in the GeMn nanocolumns with an exchange field $|H_E| = 100$ Oe and a coercive field enhancement of about 70 Oe (Fig.5) with respect to the ZFC value (650 Oe). A similarly enhanced EB at low-cooling fields has been reported by Pan et al. in cobalt oxide [16]. This effect was ascribed to the relative orientation between the cooling field and measurement field and the (intrinsic) easy magnetization axis of the system.

The exchange field obtained in our GeMn-oxide/GeMn nanocolumns is comparable to the exchange field values previously reported in other ferromagnetic semiconductor-based heterostructures: 90-180 Oe in MnO/(Ga,Mn)As and IrMn /(Ga,Mn)As [17, 18, 19] and around 70 Oe in MnO/Cr-doped GaN [20].

A MnO(3nm)/GeMn(40nm) bi-layer has been prepared by naturally oxidizing in ambient atmosphere a pure Mn film deposited on GeMn nanocolumns. An exchange field $H_E$ (MnO/GeMn) = -70 Oe and $\Delta H_C$ (MnO/GeMn) = 410 Oe was obtained in this case upon a similar FC procedure under a positive field. Annealing in air at 200 °C for 150 s induces an additional coercive field increase of 660 Oe and a shift of the magnetization curve of -150 Oe. This enhanced EB is explained by the presence of a more defined antiferromagnetic Mn-O compound in the system and by the thicker antiferromagnetic layer (MnO + oxidized-GeMn caps) that is obtained upon annealing in air. This result supports also that MnO is indeed at the origin of the observed EB in GeMn-oxide caps/GeMn nanocolumns.

In conclusion, exchange biasing of self-organized ferromagnetic GeMn nanocolumns by oxidized-GeMn caps is reported. The XAS measurements performed in GeMn-oxides shows a multiplet structure that is typical of the $Mn^{2+}$ valence state in MnO. The magnetic measurements show a clear exchange bias revealed by the hysteresis shift toward negative magnetic fields upon positive FC (and vice versa) that is accompanied by the enhancement of the coercive field. The observed exchange bias is attributed to the exchange coupling at the interface between the ferromagnetic GeMn nanocolumns and the antiferromagnetic MnO-like caps.

## ACKNOWLEDGEMENTS

This work has been supported by the French national research agency (ANR) under project N° **ANR**-07-NANO-003-02.

## REFERENCES

1. N. Spaldin and M. Fiebig, Science 309, 391 (2005).

2. D. D. Awschalom and M. E. Flatté, Nat. Phys. 3, 153 (2007).

3. D. Chiba, M. Sawicki, Y. Nishitani, Y. Nakatani, F. Matsukura and H. Ohno, Nature 455, 515 (2008).

4. K. F. Eid, M. B. Stone, K. C. Ku, O. Maksimov, P. Schiffer, N. Samarth, T. C. Shih and C. J. Palmstrøm, Appl. Phys. Lett. 85, 1556 (2004).

5. Z. Wei, A. Sharma, A. S. Nunez, P. M. Haney, R. A. Duine, J. Bass, A. H. MacDonald and M. Tsoi, Phys. Rev. Lett. 98, 116603 (2007).

6. X. Chen, A. Hochstrat, P. Borisov, and W. Kleemann, Appl. Phys. Lett. 89, 202508 (2006).

7. V. Laukhin, V. Skumryev, X. Marti, D. Hrabovsky, F. Sanchez, M.V. Garcia-Cuenca, C. Ferrater, M. Varela, U. Lüders, J. F. Bobo, and J. Fontcuberta, Phys. Rev. Lett. 97, 227201 (2006).

8. M. Jamet, A. Barski, T. Devillers, V. Poydenot, R. Dujardin, P. Bayle-Guillemaud, J. Rothman, E. Bellet-Amalric, A. Marty, J. Cibert, R. Mattana, and S. Tatarenko, Nat. Mater. 5, 653 (2006).

9. T. Devillers, M. Jamet, A. Barski, V. Poydenot, P. Bayle-Guillemaud, E. Bellet-Amalric, S. Cherifi, and J. Cibert, Phys. Rev. B 76, 205306 (2007).

10. T. Fukushima, K. Sato, H. Katayama-Yoshida and P. H. Dederichs, Jap. J. Appl. Phys. 45 L416 (2006).

11. G.V. Samsonov, The Oxide Handbook, Plenum Press, New York (1973).

12. A. Continenza and G. Profeta Phys. Rev. B 78, 0852151 (2008).

13. B. Gilbert, B. H. Frazer, A. Belz, P. G. Conrad, K. H. Nealson, D. Haskel, J. C. Lang, G. Srajer, and G. De Stasio, J. Phys. Chem. A 2003, 107, 2839-2847

14. S. Andrieu, E. Foy, H. Fischer and M. Alnot, F. Chevrier, G. Krill and M. Piecuch, Phys. Rev. B 58, 8210 (1998)

15. P. Gambardella, L. Claude, S. Rusponi, K. J. Franke, H. Brune, J. Raabe, F. Nolting, P. Bencok, A. T. Hanbicki, B. T. Jonker, C. Grazioli, M. Veronese, and C. Carbone, Phys. Rev. B 75, 125211 (2007).

16. M. Pan, B. Youa , Y. Zhao, S. Wang, M. Lu, A. Hu, H. Zhai, S. Zhou, Physics Letters A 313, 442 (2003).

17. K. F. Eid, M. B. Stone, O. Maksimov, T. C. Shih, K. C. Ku, W. Fadgen, C. J. Palmstrøm, P. Schiffer, and N. Samarth, J. Appl. Phys. 97, 10D304 (2005)

18. H. T. Lin, Y. F. Chen, P.W. Huang, S. H. Wang, J. H. Huang, C.H. Lai, W. N. Lee, and T. S. Chin, Appl. Phys. Lett. 89, 262502 (2006)

19. Z. Ge, W. L. Lim, S. Shen, Y. Y. Zhou, X. Liu, J. K. Furdyna, and M. Dobrowolska, Phys. Rev. B 75, 014407 (2007).

20. H. X. Liu, Stephen Y. Wu, R. K. Singh, and N. Newman, J. Appl. Phys. 98, 046106 (2005)

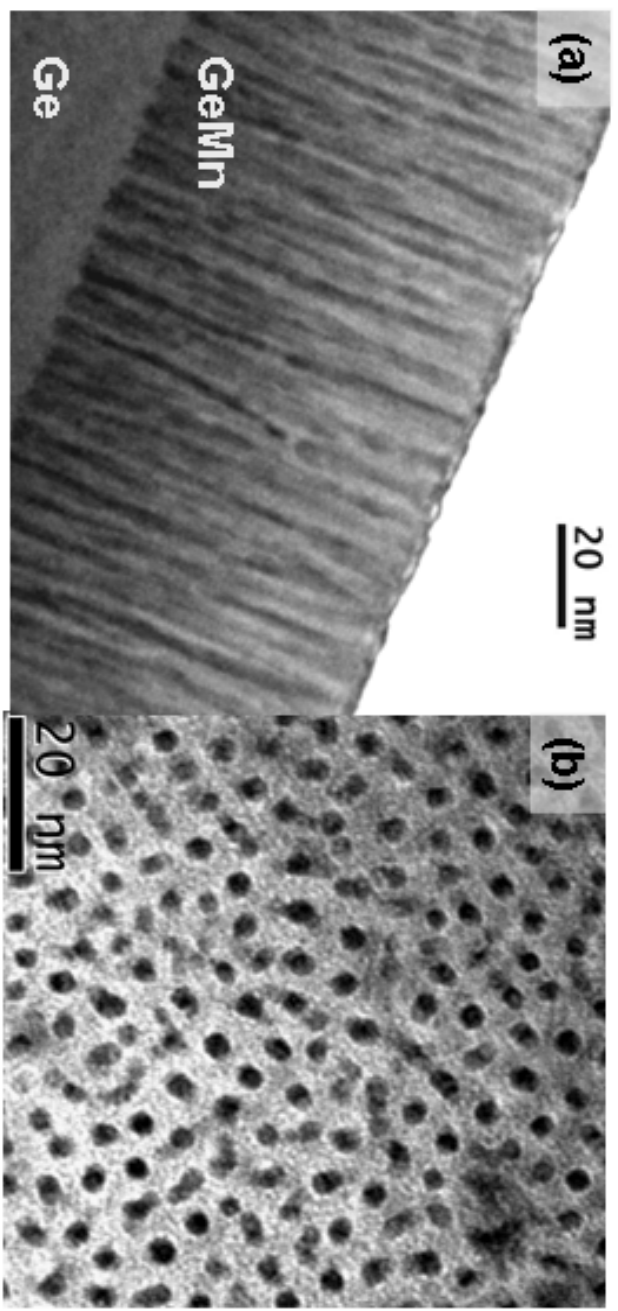

**Fig. 1** TEM observations in cross-section (a) and plane view (b) of a GeMn film containing 10% of Mn and grown at about 100°C on Ge(001). We can clearly observe self-assembled nanocolumns.

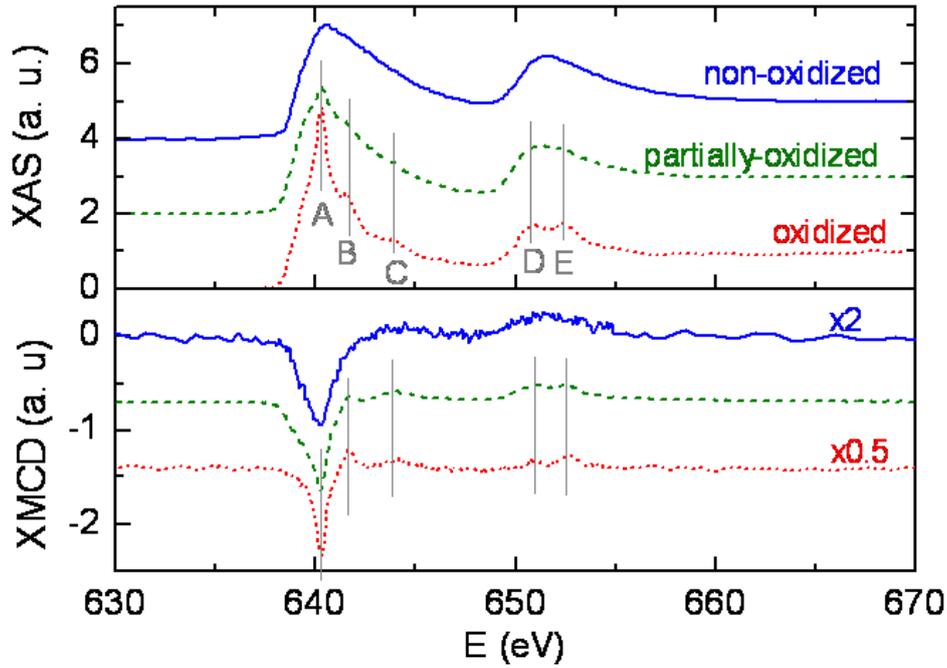

**Fig. 2** Isotropic XAS (top) and the corresponding XMCD (bottom) measured at 5 K and 5 T in: non-oxidized GeMn nanocolumns (solid line), GeMn-oxide(5 nm)/GeMn(75 nm) (dashed line) and GeMn-oxide(20 nm)/GeMn(60 nm) (dotted line). The spectra have been vertically shifted. The spectra obtained in non-oxidized and oxidized samples have been measured with two different experimental setups at BESSYII and ESRF storage rings, respectively. Therefore, no information about the energy shift after oxidation could be extracted from the measurements.

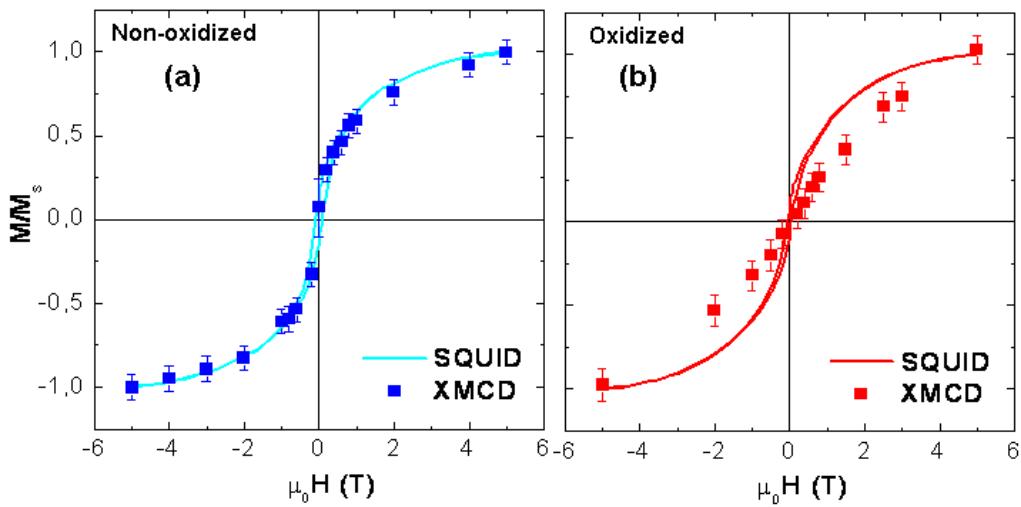

**Fig. 3**. Magnetization curves measured by using SQUID (solid line) and XMCD (circles) in: (a) non-oxidized GeMn (80 nm) and (b) GeMn-oxide(20 nm)/GeMn(60 nm) measured at 5 K and normalized to the magnetization value measured at 5 T. The XMCD signal is the magnetic field-dependent amplitude of the dichroism at the Mn $L_3$ edge.

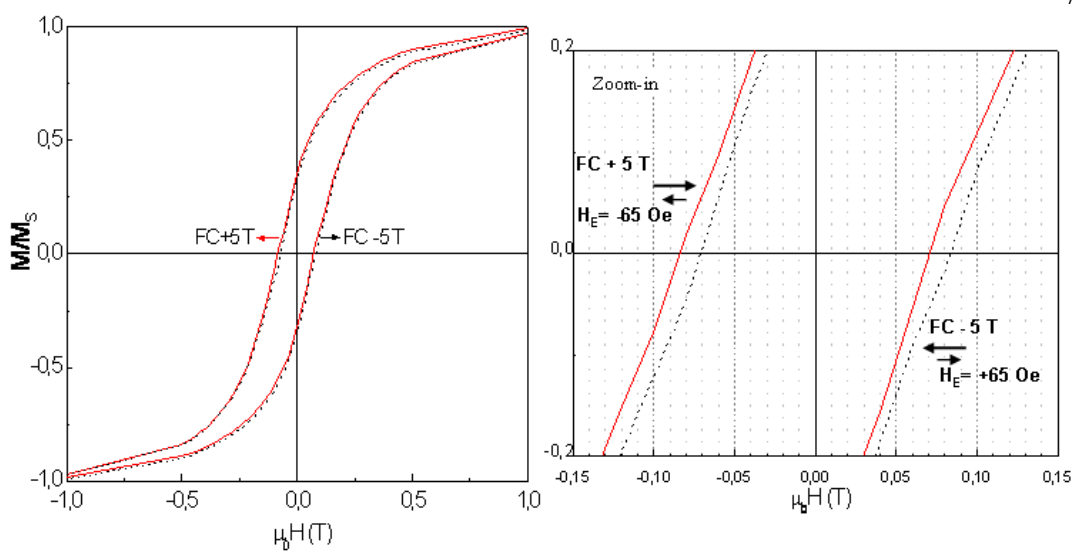

**FIG. 4** Magnetization curves after cooling the sample from room temperature to 5 K under a magnetic field of 5 T (solid line) and -5 T (dashed line). The loops are symmetrically shifted by $H_E$ = 65 Oe in the direction opposite to the cooling field while the coercive field increases by $\Delta H_C$ ~ 60 Oe with respect to the zero-field cooled value ($H_C$(ZFC) = 650 Oe).

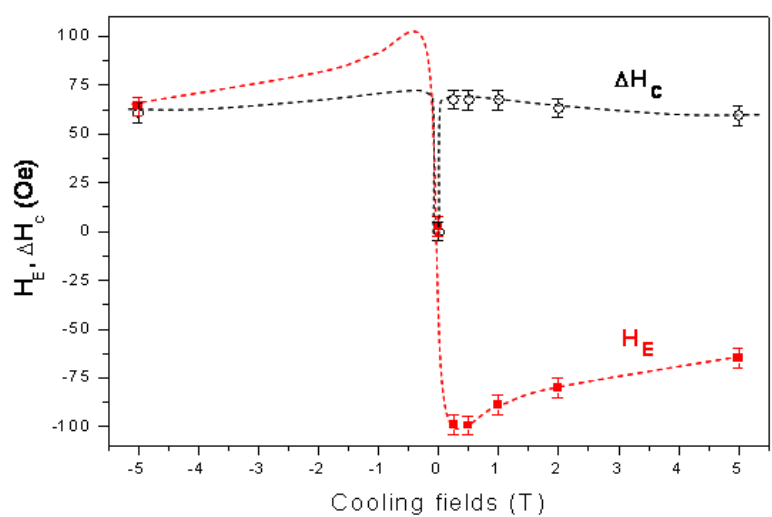

**Fig. 5** Variation of the exchange field $H_E$ (squares) and the coercive field variation $\Delta H_C$ (open circles) as a function of the cooling field. All the measurements were performed at 5 K. The dashed lines are guides of the eyes.